\input{aipcheck}

\documentclass[
    ,final            
  ]
  {aipproc}

\layoutstyle{6x9}

\begin{document}

\title{Periodic morphological changes\\in gamma-ray binaries}
\classification{97.80.Jp; 97.60.Gb; 97.20.Ec; 98.70.Rz; 98.38.Fs}
\keywords      {radio continuum: stars -- binaries: close -- gamma rays: stars -- X-rays: binaries -- radiation mechanism: non-thermal}

\author{Javier Mold\'on}{
  address={Departament d'Astronomia i Meteorologia, Institut de Ci\`encies del
Cosmos (ICC), Universitat de Barcelona (IEEC-UB), Mart\'{\i} i Franqu\`es 1,
08028 Barcelona, Spain}
}

\author{Marc Rib\'o}{
  address={Departament d'Astronomia i Meteorologia, Institut de Ci\`encies del
Cosmos (ICC), Universitat de Barcelona (IEEC-UB), Mart\'{\i} i Franqu\`es 1,
08028 Barcelona, Spain}
}

\author{Josep M. Paredes}{
  address={Departament d'Astronomia i Meteorologia, Institut de Ci\`encies del
Cosmos (ICC), Universitat de Barcelona (IEEC-UB), Mart\'{\i} i Franqu\`es 1,
08028 Barcelona, Spain}
}

\begin{abstract}
Gamma-ray binaries allow us to study physical processes such as particle acceleration up to very-high energies and gamma-ray emission and absorption with changing geometrical configurations on a periodic basis. They produce outflows of radio-emitting particles whose structure can be imaged with Very Long Baseline Interferometry (VLBI). We present recent and new VLBI observations of PSR~B1259$-$63, LS 5039, LS I +61 303, and HESS~J0632+057. For the first three cases the results show the repeatability of their radio structures with the orbit of the binary system.
\end{abstract}

\maketitle


\section{Introduction}

Some galactic binary systems have been detected in high-energy (HE; $>100$~MeV) and/or very-high-energy (VHE; $>100$~GeV) gamma rays (see \cite{paredes11} and references therein). Among these systems, gamma-ray binaries show a broadband non-thermal spectral energy distribution from radio to gamma-rays that is dominated by MeV--GeV photons \cite{dubus06}. The spectral and brightness properties appear to be synchronised with the orbit of the binary system and they do not show evidence of the presence of an accretion disc. Five binary systems have been classified as gamma-ray binaries: PSR~B1259$-$63, LS~5039, LS~I~+61~303, HESS~J0632+057, and 1FGL~J1018.6$-$5856 \citep{paredes11}. The latter has not been clearly detected at TeV energies yet. In all of them the optical companion is a young massive star. The VHE gamma-ray emission can be interpreted as the result of inverse Compton upscattering of stellar UV photons by relativistic electrons, although hadronic models do exist as well. The acceleration of electrons can be explained by two excluding scenarios: acceleration in the jet of a microquasar powered by accretion \cite{bosch-ramon06}, or shocks between the relativistic wind of a young non-accreting pulsar and the wind of the stellar companion \citep{maraschi81, dubus06, khangulyan07}. Gamma-ray binaries allow us to study physical processes such as particle acceleration up to very-high energies and gamma-ray emission and absorption with changing geometrical configurations on a periodic basis.

Gamma-ray binaries are known to display non-thermal synchrotron radio emission of the order of 0.1--100~milliJansky (mJy). Relativistic electrons that travel away from the system can produce radio emission up to distances of several AU. Therefore, the radio outflows can easily reach projected angular separations of the order of milliarcseconds (mas), which are directly observable by means of Very Long Baseline Interferometry (VLBI) at radio wavelengths.

\section{Extended radio emission in gamma-ray binaries}

Four gamma-ray binaries have been observed with VLBI. We describe the main properties of the systems and the results from our recent and new VLBI observations.

\paragraph{PSR~B1259$-$63}

The binary system PSR~B1259$-$63/LS~2883 is formed by a young 48~ms radio pulsar in an eccentric orbit of 3.4~years around a massive main-sequence star \citep{johnston94}. The spectral type of the massive star, O9.5\,Ve, and some of the binary parameters have been recently updated by \cite{negueruela11}, who obtained a distance to the system of $2.3\pm0.4$~kpc. PSR~B1259$-$63 is the only gamma-ray binary in which radio pulsations have been detected up to now.

We observed PSR~B1259$-$63 with the Australian Long Baseline Array (LBA) at 2.3~GHz in 2007 during three epochs corresponding to 1.3, 21.2, and 315.5 days after the periastron passage \citep{moldon11_psr}. In the first two epochs the radio source showed extended emission up to projected distances of $\sim50$~mas, or $\sim120$~AU. The third epoch, obtained closer to the apastron passage, showed a faint point-like source, which was interpreted as emission from the pulsar that indicates the position of the binary system. The position of the binary system within the nebula for the first two epochs was estimated using the size of the orbit and the proper motion of the system. The results showed that the peak of the radio emission is displaced from the binary system. Preliminary VLBI images of the 2010 periastron passage of PSR~B1259$-$63 suggest that its VLBI structure varies periodically (Mold\'on et al., in preparation).


\paragraph{LS~5039}

This system, located at $2.9\pm0.8$~kpc, is composed by an ON6.5\,V((f)) massive star and a compact object of unknown mass in a 3.9~d orbit \citep{casares12_ls}.  
The radio emission appears extended when observed with VLBI on mas scales. At 5~GHz, the source shows a main core and extended bipolar emission that has been detected at projected angular distances between 1 and 180~mas (3--500~AU) from the core \citep{paredes00,paredes02,ribo08}.

We observed LS~5039 with the VLBA at 5~GHz during five consecutive days, covering a whole orbital cycle \citep{moldon12_vlbi}. The radio morphology on mas scales is variable, with changes in less than 24 hours. Images at similar orbital phases show a similar morphology. We found that this behaviour is stable on time scales of years. Therefore, the gamma-ray binary LS~5039 shows periodic orbital morphological variability. We computed a model of the evolution of an outflow of relativistic electrons accelerated as a consecuence of the interaction of the massive star and the wind of a young non-accreting pulsar. The model accounts for the main morphological features of LS 5039 \citep{moldon12_vlbi}.

\paragraph{LS~I~+61~303}

This gamma-ray binary contains a rapidly rotating B0e main sequence star with a stable equatorial shell, and a compact object orbiting it every 26.5~d in an eccentric orbit with $e=0.72$ \citep{casares05_lsi}. It displays quasi-periodic radio outbursts at radio, X-rays and gamma-rays modulated by a long-term periodicity of $\sim4.6$~yr \citep{gregory02}.
\cite{dhawan06} conducted several VLBI observations showing an orbital variability that these authors interpreted as the signature of a cometary tail produced in the colliding winds scenario. Another interpretation of the data suggests that the changes are compatible with a precessing microblazar \citep{massi12}. 

We observed LS~I~+61~303 with the VLBA in 2007 at 8.4 and 2.3~GHz during five epochs separated by two days and covering an X-ray/TeV outburst. We found extended and variable emission on scales of 5--10~AU. The observed morphology is compatible with the one found in observations conducted one year before at similar orbital phases, which suggests that the periodic orbital modulation is stable on time scales of years. We obtained accurate astrometry at two radio frequencies and found the relative offsets between them. The behaviour found is similar to the one expected from an outflow whose axis passes close to the line of sight of the observer (Mold\'on et al., in preparation).

\paragraph{HESS~J0632+057}

The proposed optical counterpart of HESS~J0632+057 is the massive B0pe star MWC~148 \citep{hinton09}. A periodicity of $321\pm5$~d was revealed by long-term X-ray observations conducted with {\it Swift}/XRT \citep{bongiorno11}. MWC~148 was confirmed to be a binary system by means of radial velocity observations \citep{casares12_j0632}. The radio counterpart is variable, although no detailed radio light curve is available.

We observed HESS J0632+057 with the EVN at 1.6~GHz in ToO mode in two epochs: February 15 and March 17, 2011, following the report of an X-ray outburst. The first observation (e-EVN) was conducted 9 days after the peak of the X-ray outburst, whereas the second observation (full EVN) was conducted 30 days later. The source is point-like during run~A and displays extended emission up to projected distances of 75 AU in run~B \citep{moldon11_j0632}. The peak of the emission is displaced 21 AU between runs. The radio counterpart of HESS~J0632+057 is unambiguously identified with MWC~148.

\section{Conclusions}

The four gamma-ray binaries detected at TeV energies show extended structures when observed with VLBI. The common properties of the radio structures from PSR~B1259$-$63, LS~5039, and LS~I~+61~303 can be summarized in three points, which are model independent: (1) They display synchrotron radio emission that forms structures on AU scales. (2) These radio structures show periodic orbital morphological variability, although erratic changes might occur. (3) The position of the peak of the emission suffers displacements significantly larger than the size of the orbit. For HESS~J0632+057, the second property has not been verified yet due to the lack of observations during different orbital cycles. In Table~\ref{table1} we show the list of known gamma-ray binaries with the orbital period of the system, and the properties of the VLBI structure of each source. With the current data we can see that most of them display periodic morphological changes, coupled with the orbital period, and therefore this behaviour is expected for all gamma-ray binaries.

\begin{table} 
\caption{Summary of the VLBI observations of the known gamma-ray binaries.}
\label{table1} 
\begin{tabular}{l r@{.}l  cc cc}
\hline
Source              & \multicolumn{2}{c}{$P_{\rm orb}$ [d]}     & Observed    & VLBI structure   & Variability       \\
\hline
LS~5039              & 3&9   &  Yes   & Yes &  Periodic changes         \\
1FGL~J1018.6$-$5856  & 16&6   &  No\tablenote{The source was very recently observed with the Australian LBA (PI: P. Edwards).}   & ? &  --        \\
LS~I~+61~303         & 26&5   &  Yes  & Yes &  Periodic changes     \\
HESS~J0632+057       & \multicolumn{2}{c}{321~~}   &  Yes  & Yes &  Variable      \\
PSR~B1259$-$63       & 1236&8   &  Yes  & Yes &  Periodic changes      \\
\hline
\end{tabular}

\end{table}


\begin{theacknowledgments}
J.M., M.R., and J.M.P. acknowledge support by MINECO under grants AYA2010-21782-C03-01 and FPA2010-22056-C06-02. J.M. acknowledges support by MINECO under grant BES-2008-004564. M.R. acknowledges financial support from MINECO and European Social Funds through a \emph{Ram\'on y Cajal} fellowship. J.M.P. acknowledges financial support from ICREA Academia. 
\end{theacknowledgments}



\bibliographystyle{aipproc}   

\bibliography{proc}

\IfFileExists{\jobname.bbl}{}
 {\typeout{}
  \typeout{******************************************}
  \typeout{** Please run "bibtex \jobname" to optain}
  \typeout{** the bibliography and then re-run LaTeX}
  \typeout{** twice to fix the references!}
  \typeout{******************************************}
  \typeout{}
 }

\end{document}